\documentclass[12pt]{article}

\usepackage{amsmath}
\usepackage{amssymb}
\usepackage{graphicx}
\usepackage{epsfig}
\usepackage{multirow}
\usepackage{dcolumn}
\usepackage{bm}
\usepackage{cite}
\usepackage[small]{caption}
\usepackage[usenames]{color}

\def\wt{\widetilde}
\def\sla#1{\setbox0=\hbox{$#1$}\dimen0=\wd0
      \setbox1=\hbox{/} \dimen1=\wd1 \ifdim\dimen0>\dimen1
      \rlap{\hbox to \dimen0{\hfil/\hfil}} #1
      \else
      \rlap{\hbox to \dimen1{\hfil$#1$\hfil}}
      /   \fi}
\def\MET{\sla{E}_{\rm T}}
\def\MPT{\sla{p}_{\rm T}}

\def\lsim{\raise0.3ex\hbox{$\;<$\kern-0.75em\raise-1.1ex\hbox{$\sim\;$}}}
\def\gsim{\raise0.3ex\hbox{$\;>$\kern-0.75em\raise-1.1ex\hbox{$\sim\;$}}}
\topmargin
-1.5cm
\textwidth
15.05cm
\textheight
23.5cm
\oddsidemargin
0.7cm
\evensidemargin
0.7cm

\def\Sn1{{S^0}^*}
\def\Pn1{{P^0}^*}
\def\Cn1{{S^\pm}^*}

\def    \beq            {\begin{equation}}
\def    \eeq            {\end{equation}}
\def    \bea           {\begin{eqnarray}}
\def    \eea           {\end{eqnarray}}

\def\fbi{{\rm fb}^{-1}}   
\def\la{\lambda} 
\def\ka{\kappa}

\def \wt {\widetilde}
\def\h{h^0_}
\def\a{a^0_}

\def\mt{m_{\tau}}

\def\bmumu{B^0_s\to\mu^+\mu^-}
\def\bsg{b\to s \gamma}
\def\ps{\displaystyle{\not} {\rm p}}

\begin{document}
\def\thefootnote{\fnsymbol{footnote}}

\begin{flushright}
IFT-UAM/CSIC-13-002\\
FTUAM-13-124\\
CERN-PH-TH/2012-366
\end{flushright}

\begin{center}
  {\bf {\Large Probing the two light Higgs scenario \\
      \vskip 0.1in
      in the NMSSM with a low-mass pseudoscalar}}
\end{center}

\begin{center}{\large
    David~G.~Cerde\~no$^{a,b,}$\footnote{Email: davidg.cerdeno@uam.es},
    Pradipta~Ghosh$^{a,b,}$\footnote{Email: pradipta.ghosh@uam.es}, and
    Chan~Beom~Park$^{c,}$\footnote{Email: chanbeom.park@cern.ch}
  }
\end{center}

\begin{center}
$^a${\em Departamento de F\'{\i}sica Te\'{o}rica, Universidad
Aut\'{o}noma de Madrid,\newline
Cantoblanco, E-28049 Madrid, Spain}\\
$^b${\em Instituto de F\'{\i}sica Te\'{o}rica UAM/CSIC, Universidad
Aut\'{o}noma de Madrid,\newline
Cantoblanco, E-28049 Madrid, Spain}\\
$^c${\em TH Division, Physics Department, CERN, CH-1211 Geneva 23,
Switzerland}
\end{center}
\date{\today}

\vspace*{0.5cm}
\centerline{\em{\bf Abstract}}
\noindent
In this article we propose a simultaneous collider search strategy
for a pair of scalar bosons in the NMSSM through the decays of a
very light pseudoscalar. The massive scalar
has a mass around $126$ GeV while the lighter one can have
a mass in the vicinity of $98$ GeV (thus explaining
an apparent LEP excess) or be much lighter.
The successive decay of this scalar pair into two
light pseudoscalars, followed by leptonic pseudoscalar
decays, produces clean multi-lepton final states with
small or no missing energy. Furthermore, this analysis
offers an alternate leptonic probe for the $126$~GeV
scalar that can be comparable with the $ZZ^\ast$ search channel.
We emphasize that a dedicated experimental search 
for multi-lepton final states can be
an useful probe for this scenario and, in general, for the NMSSM Higgs sector.
We illustrate our analysis with two representative benchmark points and show how
the LHC configuration with $8$ TeV center-of-mass energy and $25~\fbi$
of integrated luminosity can start testing this scenario,
providing a good determination of the light pseudoscalar mass and a relatively
good estimation of the lightest scalar mass.


\newpage
\renewcommand{\thefootnote}{\arabic{footnote}}
\setcounter{footnote}{0}

\noindent
\section{Introduction}\label{intro}
The experimental evidence for a scalar boson with a mass around 
126~GeV~\cite{:2012gk,CMS:2012gu} has been received with enthusiasm 
by the particle physics community, since it could correspond to the 
missing piece of the standard model (SM), the Higgs boson.
Although most of the properties of this new particle (in particular 
the branching ratios of various decay modes) are compatible with a 
SM-like Higgs, an apparent di-photon excess~\cite{diphoton-ATLAS,diphoton-CMS}, 
has been considered as a possible hint for physics beyond the SM, such as supersymmetry (SUSY).
However, this excess remains yet to be confirmed~\cite{Baglio:2012et}.

It has been pointed out that in order to reproduce the mass of the Higgs boson 
in the Minimal Supersymmetric Standard Model (MSSM) one generally needs a very 
heavy spectrum, raising concerns about the naturalness of this construction.
This problem can be alleviated in extended Supersymmetric constructions. 
In particular, the Next-to-Minimal Supersymmetric Standard Model
(NMSSM) can accommodate a 126 GeV
scalar \cite{Ellwanger:2011aa,Gunion:2012zd,Cao:2012yn,
Gunion:2012he,Belanger:2012tt,King:2012tr}, while maintaining the
idea of naturalness \cite{Ellwanger:2011mu,Arvanitaki:2011ck,Kang:2012sy,King:2012tr}.
The advantage of the NMSSM resides in the presence of an extra singlet field, which provides
an additional contribution to the tree-level Higgs
mass~\cite{Drees:1988fc,Ellis:1988er,Binetruy:1991mk}.
Thus, contrary to what happens in the MSSM, loop
corrections can be smaller and the correct Higgs mass 
is achieved for a lighter supersymmetric spectrum.
This extra singlet superfield $\hat S$ is introduced to promote
the bilinear $\mu$-term $\mu\hat H_d \hat H_u$ in the MSSM
to a trilinear coupling $\lambda \hat S\hat H_d \hat H_u$.
After electroweak symmetry breaking (EWSB) takes place, the vacuum expectation
value (VEV) of the singlet field triggers an effective $\mu$
parameter, $\lambda v_s$, which is naturally of the order of the electroweak 
scale, thus curing the $\mu$-problem of the MSSM~\cite{Kim:1983dt}. 
On top of this, the singlet-doublet mixing of the resulting scalar states is
useful to produce the observed di-photon excess \cite{Ellwanger:2010nf,Ellwanger:2011aa}.

The new singlet superfield induces one extra CP-even and CP-odd
neutral scalars as well as one additional neutral fermion, thereby
leading to a very rich phenomenology.
In particular, a light pseudoscalar Higgs is viable if its singlet 
component is large~\cite{Gunion:1996fb,Dobrescu:2000jt,Ellwanger:2003jt,
Dermisek:2005ar,Dermisek:2010mg,Ellwanger:2005uu,Dermisek:2008uu,
Dermisek:2009fd}. Similarly, the lightest supersymmetric particle
(LSP) can also be singlino-like \cite{Abel:1992ts} or, in general, a mixed 
state with interesting implications for dark matter searches  
(see, e.g., Refs.\,\cite{Cerdeno:2004xw,Cerdeno:2007sn}).

The presence of light scalar and pseudoscalar Higges can lead to characteristic
Higgs-to-Higgs cascade decays~\cite{Gunion:1996fb,
Dobrescu:2000jt}. Although these cascades
are generally suppressed, a substantial
enhancement can occur when particles are light
and therefore induce distinctive signatures.\footnote{
To some extent, a scenario like this is a good illustration of the
``no-lose'' theorem for the NMSSM~\cite{Gunion:1996fb,Dobrescu:2000jt,
  Ellwanger:2001iw, Ellwanger:2003jt,Miller:2004uh,Ellwanger:2005uu,
  Dermisek:2005ar, Chang:2006bw,Forshaw:2007ra, Belyaev:2008gj}.
}
For example, if the lightest pseudoscalar, denoted as $a^0_1$,
has a mass $m_{a^0_1}\gsim 2 m_b$, the process
$\h1\to 2\a1\to 2b \bar{b}$ with $4b$ final states is dominant
\cite{Ellwanger:2003jt,Abdallah:2004wy,Dermisek:2007yt,
Carena:2007jk,Almarashi:2011hj,Almarashi:2011bf}.
This seems a good channel for collider study,\footnote{
Another possible search channel is $2b2\tau$.}
however, in hadron colliders like the LHC this signature 
is hidden by a huge QCD background. A viable alternative
is looking for $4\tau$ \cite{Gunion:1996fb,
Ellwanger:2001iw,Abbiendi:2002in,Ellwanger:2003jt,
Belyaev:2008gj,Dermisek:2008uu,Lisanti:2009uy,
Almarashi:2011hj} or $2\mu2\tau$ \cite{Lisanti:2009uy}
final states, for which the problem of large QCD backgrounds is somehow
ameliorated. However, this approach suffers
from the poor tau-identification efficiency, dependent
on the transverse momentum \cite{Bayatian:2006zz}. Besides, since tau 
decays are always accompanied by neutrinos, there is no sharp
peak of $\a1$ invariant mass. For these reasons, as discussed in
the literature, $4\mu$ \cite{Zhu:2006zv,Belyaev:2010ka,Almarashi:2011hj},
$4\gamma$ \cite{Dobrescu:2000jt,Chang:2006bw}
or $2\gamma+2$-jets \cite{Martin:2007dx} search channels with
$m_{\a1}<2\mt$ turn out to be the ideal way
to probe NMSSM Higgses with very light pseudoscalars.

Recently, the possibility of having two light scalar Higgses 
in NMSSM has received a lot of attention.
One of these Higgses, $\h2$, would correspond to the 
scalar observed by ATLAS and CMS
with a mass around 126~GeV and the other one, $\h1$, with a mass around 98
GeV~\cite{Belanger:2012tt}, would be consistent with the small excess in 
the LEP search for $e^+e^-\to Z h$, $h\to b \bar{b}$
\cite{Barate:2003sz}.\footnote{A two light Higgs but without a low-mass 
pseudoscalar has been discussed in the MSSM \cite{Heinemeyer:2011aa},
 as well as in the NMSSM \cite{Kang:2013rj}. }
Alternative scenarios with an even lighter $\h1$ have also been
considered earlier \cite{Ellwanger:2003jt,Djouadi:2008uw}.
Since the recent LHC observation of the 126 GeV
scalar, which is well compatible with a SM-like Higgs boson, many
of these studies have become extremely constrained, although a small window
for new physics is still open if the di-photon excess is confirmed.
On the other hand, scenarios with very light pseudoscalars are 
also affected by stringent experimental bounds.
This is the case for $m_{\a1}<2m_b$ (which leads to
leptonic final states $4\mu$, $2\mu2\tau$, and $4\tau$) \cite{Forshaw:2007ra,
Abazov:2009yi,Lisanti:2009uy,Dermisek:2009fd}, with tight constraints in the case
$m_{\a1}<2m_\tau$ for CP-even Higgs
masses in the range of 86 -- 150 GeV by the recent CMS
analysis~\cite{Chatrchyan:2012cg}, especially in the case of singlet-like states.
The latter is a consequence of the fact that BR$(\h1/\h2 \to\a1\a1)\sim 1$
and thus $\sigma(pp\to \h1/\h2 \to 2\a1\to 4\mu)$ is sizable and would
have already been observed.\footnote{
One way to avoid this is to consider the case when
the production cross section for a singlet-like $\h1$/$\h2$ is
very suppressed.}

In this work we investigate potential detection channels for scenarios involving 
very light pseudoscalar particles and two light scalar Higgses in the NMSSM.
For concreteness, we consider a SM-like scalar Higgs in the range 
$124~{\rm GeV} <m_{\h2}<128$ GeV (consistent
with LHC findings) and a lighter $\h1$ in the mass range 96 -- 100 GeV 
(to account for the LEP excess) or lighter than 86 GeV (to avoid the current CMS limit).
In order to avoid the above mentioned constraints that affect the lightest 
pseudoscalar, $\a1$ is assumed to have a mass in the range $2m_\tau \lesssim m_{\a1} < 2m_b$,
since the $\mu^+\mu^-$ final state
is then a sub-leading leptonic mode with a branching ratio
of the order of $10^{-2}$ and $\sigma(pp\to\h1\to 2\a1\to 4\mu)$
is small enough to elude the CMS limit of the $4\mu$ search.
We point out that the SM-like Higgs decay modes $\h2\to 2\a1\to 4\mu$, $2\mu2\tau$, and $4\tau$
have branching fractions comparable to those for the same final states from
$\h2\to ZZ^\ast$ and therefore provide alternative four-lepton signatures 
accompanied with zero or a small missing energy at the LHC.
We also note that the long Higgs-to-Higgs cascades of $\h2\to\h1\h1\to
4\a1$ can be open if $m_{\h1} < m_{\h2}/2$, which constitutes a clean and 
distinctive multi-lepton signal at the LHC.
We design a set of cuts that isolates the signal events from the background,
study the distributions of the kinematic variable $M_{\rm T2}$~\cite{Lester:1999tx}, 
and determine to what extent the masses of the pseudoscalar and scalar Higgs particles 
can be reconstructed. In our analysis we include constraints from $B$-physics
\cite{Asner:2010qj,Chatrchyan:2012rg,Aaij:2012nna,Amhis:2012bh},
the muon anomalous magnetic moment \cite{Bennett:2006fi,Jegerlehner:2009ry,
Davier:2010nc,Hagiwara:2011af}, and the LHC result for $\h2\to\gamma\gamma$.
In addition, we impose the WMAP bound on the lightest neutralino relic abundance
\cite{Komatsu:2010fb} and we ensure that its spin-dependent scattering 
cross section is consistent with XENON100 data \cite{Aprile:2012nq}.

The paper is organized as follows. We start with a brief
description of the model and determine the relevant signatures
in Sec.\,\ref{model}. The choice of model parameters is justified in 
Sec.\,\ref{analysis} where we also describe the details of the analysis.
Sec.\,\ref{results} is dedicated to the analysis
of the numerical results. We also introduce some discriminating variables to
suppress SM backgrounds and we investigate the statistical significance of the
proposed signals. Finally we summarize our conclusions in Sec.\,\ref{conl}.

\noindent
\section{Higgs signals with a low-mass pseudoscalar in the NMSSM}\label{model}
The superpotential
of the $\mathbb{Z}_3$-symmetric NMSSM
is given by
%
\beq\label{superpotential}
W = W^\prime -\epsilon_{ab} \lambda \hat S \hat H^a_d\hat H^b_u
+\frac{1}{3}\kappa \hat S \hat S \hat S,
\eeq
where $W^\prime$ is the MSSM superpotential without the $\mu$-term,
$\hat S$ is a superfield singlet under the SM gauge group, and $a,\,b
= 1,\,2$ are SU(2)$_L$ indices. Similarly,  
with $\mathcal{L}'_{\text{soft}}$ representing the MSSM soft-terms
without the $B_\mu$-term, the Lagrangian 
containing the soft SUSY-breaking terms
in a supergravity framework is given by
\bea\label{soft}
-\mathcal{L}_{\text{soft}} =
-\mathcal{L}'_{\text{soft}}
+ m_{S}^2 {\wt S^\ast}  {\wt S}
-\epsilon_{ab} (\lambda A_{\lambda}) \wt S H_d^a  H_u^b+
\frac{1}{3} (\kappa A_{\kappa}) \wt S \wt S \wt S + \text{H.c.}\,.
\label{Lsoft}
\eea
%
The last trilinear term in Eq.\,(\ref{superpotential}) is
essential to avoid an unacceptable axion associated with the
breaking of a global U(1) symmetry \cite{Ellis:1988er}.
After EWSB, this term also generates an effective Majorana mass
$2\kappa v_s$ for the singlino-like neutralino.
The second term of Eq.\,(\ref{superpotential}) not only
generates an effective $\mu$ parameter, $\mu_{\rm eff}=\la v_s$, but also
provides an extra tree-level contribution to the lightest doublet-like
Higgs boson mass. The complete expression is 
now \cite{Drees:1988fc,Ellis:1988er,Binetruy:1991mk},
%
\bea\label{tree-higgs}
m^2_{h^0} \leq M^2_Z \cos^2{2\beta} +
3.62\, M^2_Z\, {\lambda}^2 \sin^2{2\beta},
\eea
where $\tan\beta=v_u/v_d$ is the ratio of up and down-type
Higgs VEVs. Thus even maintaining perturbativity
 $(\la\lesssim 0.7)$, the extra contribution to the tree-level
Higgs mass can be sizable for small values of $\tan\beta$.
The last term of Eq.\,(\ref{Lsoft}) is important to understand
the phenomenology of singlet-like scalars. In particular,
when the lightest CP-even scalar $h^0_1$ as well as
the lightest CP-odd scalar $a^0_1$ are predominantly
singlet-like, the decay width for $h^0_1\to \a1\a1$ is 
proportional to $A_\ka\ka$.

As already mentioned in the previous section, the Higgs sector
of the NMSSM is richer than that of the MSSM, featuring three CP-even scalar
states, ($\h1$, $\h2$, $\h3$) and two pseudoscalars, ($\a1$, $\a2$).
This is of particular interest for studying Higgs cascade decays, which 
can now be long and often result in multi-particle final states
with no missing energy.
The decay chains become more interesting in the presence of light scalar 
and pseudoscalar Higgses, which can be in agreement with present constraints 
if their singlet composition is dominant.
In particular, a viable hierarchy for the scalar mass eigenstates in the 
NMSSM consists of having a light singlet-like $\h1$, a SM-like Higgs $\h2$ 
with a mass around 126~GeV and a much heavier $\h3$ which is also doublet-like.
This structure with two light scalar Higgses might produce distinctive features 
in collider searches, as compared to the signals that come from the MSSM Higgs 
sector, and therefore we concentrate on this possibility.

Although very light singlet-like pseudoscalars can give rise to
potentially characteristic signatures, one must be careful with 
recent experimental constraints. In particular, for a scalar Higgs mass 
in the mass range of 86 -- 150~GeV, the upper limit of
$\sigma(pp \to \h1/\h2 \to 2\a1) \times$BR$^2(\a1\to\mu^+\mu^-)$ has been found
to be 3 -- 5 fb, depending on the pseudoscalar mass for
$1~{\rm GeV}<m_{\a1}<2 m_{\tau}$ in the 7 TeV CMS search \cite{Chatrchyan:2012cg}.
Thus, if we want to study the 98 $+$ 126 GeV scenario for the scalar Higgs bosons,
in order to avoid this constraint, we will consider a light pseudoscalar with 
a mass in the range $2m_\tau\lesssim m_{\a1}<2m_b$. We call this Scenario I.

Another way to avoid the CMS bound is to consider a lighter scalar Higgs with $m_{\h1}<86$ GeV. Although this would relax the LHC limit on the pseudoscalar mass, in fact $2m_\tau<m_{\a1} < 2m_b$ is still very constrained from the LEP limit on $4\tau$~\cite{Beacham:2010hf} and multi-lepton/jet final states \cite{lep,Abdallah:2004wy} in the case when $m_{\h1} \lesssim m_{\h2}/2$.
We have included these constraints in our analysis and checked that viable points can still be obtained within this mass range for the lightest pseudoscalar. We call this Scenario II.

For clearness we summarise here the properties of the two kind of 
scenarios considered in our work
\begin{itemize}
 \item[(I)]
124 GeV$\lesssim m_{\h2}\lesssim$128 GeV, \newline
96 GeV$\lesssim m_{\h1}\lesssim$100 GeV, \newline
$2m_\tau \lesssim m_{\a1}<2m_b$,
\item[(II)]
124 GeV$\lesssim m_{\h2}\lesssim$128 GeV,\newline
$m_{\h1}\lesssim m_{\h2}/2$,\newline
$2m_\tau\lesssim m_{\a1}<2m_b$.
\end{itemize}
We have also checked that these benchmark scenarios are in agreement with
ATLAS and CMS direct pseudoscalar searches,
$pp\to\a1\to \mu^+\mu^-$~\cite{dimuon-ATLAS},
due to the singlet nature of $\a1$ and its dominant decay branching
fraction to $\tau$.

We are now ready to specify the potential signatures that we will study for 
these scenarios. We consider the following decay modes,
\bea\label{signal}
&&\h1\to\a1\a1\to~4\ell+\MET,
\nonumber\\
&&\h2\to\a1\a1\to~4\ell+\MET,
\eea
where $\ell=e$, $\mu$, $\tau$. We here assume that $\h1$ and $\h2$ have
been produced through gluon-fusion and consider inclusive
decay modes of the tau lepton.
The missing energy $\MET$ is associated with
neutrinos coming from tau-decays and is generally small.
It is important to emphasize that direct-pseudoscalar
decays to a muon-pair will eventually yield a clean four-muon final
state with vanishing missing energy and the only source
of electrons is leptonic $\tau$-decay, since BR$(\a1\to e^+e^-)\sim 0$.
The second decay mode can act as an alternative source of four-lepton final
state apart from $\h2\to Z Z^\ast\to4\ell$.
In the chosen corner of parameter space, we found that the predicted BR$(\h2\to
\a1\a1)$ is comparable with that of BR$(\h2\to ZZ^\ast)$, which
still matches well the SM prediction. We thus encourage our
experimental colleagues to search for possible excesses
in $4\mu$, $4\tau$ or $2\mu2\tau$ channels by relaxing the $Z$-boson
resonance condition. In the remainder of the paper we carry out a dedicated 
analysis including the discussion of SM backgrounds and
useful collider variables to search for the four-lepton 
collider signals of these decay modes.

Concerning possible SUSY backgrounds, these generally lead to
large missing energy and can therefore be distinguished with a cut on $\MET$.
There can also be sources of four-lepton final states with a small or no
missing energy, provided that one considers $R$-parity
violation~\cite{Weinberg:1981wj} in the NMSSM~\cite{Pandita:1999jd} or
$\mu\nu$SSM~\cite{LopezFogliani:2005yw}.
It is well known that $R$-parity conserving SUSY models are
characterized by large missing energy due to the production of a 
heavy neutralino LSP, which is stable and escapes the detector,
whereas this is not the case in $R$-parity violating scenarios.
Models with broken $R$-parity and non-minimal superfield content,
on the other hand, exhibit moderate or large displaced vertices
with non-prompt jets or leptons in the final states (see, for example,
Refs.~\cite{Bartl:2009an,Fidalgo:2011ky}) with more complex Higgs cascade
decays~\cite{Fidalgo:2011ky} and occasional correlations with neutrino
physics~\cite{correlations} following~\cite{Mukhopadhyaya:1998xj}, 
which can be used for discrimination.
We also note that four leptons can be produced in $\h3$ decays, 
but with a very reduced production cross section.

Notice that one distinguishing feature of Scenario II is a 
three-step cascade decay of the Higgs
boson $\h2 \to 2\h1 \to 4\a1 \to$ multi-lepton/jet.
We have checked for several points in the parameter space that, 
even when $m_{\h1}\gsim m_{\h2}/2$,
the branching ratio of $\h2\to\h1{\h1}^\ast\to4\a1$ can be as
large as $\sim 10^{-3}$ and thus multi-lepton final states
can be experimentally accessible. Due to the large number of
leptons and/or jets in the final state, the main backgrounds are 
expected to be SUSY processes (such as cascade decays
mediated by neutralinos and/or charginos), rather than SM ones.
The analysis of this decay signal will be carried out in a
separate work \cite{CGPM2} since the search strategy is 
very different than for the other two.


\noindent
\section{Experimental constraints and the choice of parameters}
\label{analysis}

The recent discovery of a 126 GeV scalar along with the reported di-photon 
excess set stringent limitations on the NMSSM parameter space. In addition,
considering bounds from low-energy observables in flavour physics,
the supersymmetric contribution to the muon anomalous magnetic moment 
$a_{\mu}^{\rm SUSY}$, and the relic abundance
of the dark matter, which in our case would correspond to the lightest
neutralino, narrow down significantly the viable regions.

In our analysis we impose compatibility with all these constraints. 
The numerical results are obtained with \textsc{nmssmtools
  3.2.1}~\cite{Ellwanger:2004xm}, which calculates the mass spectrum and 
provides predictions for low-energy observables, as well as computing the 
decay widths of Higgses~\cite{Djouadi:1991tka} and sparticles~\cite{Muhlleitner:2003vg}.
The masses of Higgs bosons include full two-loop contributions.
We also consider the lightest neutralino as a dark matter candidate 
and include the WMAP upper constraint on its relic abundance \cite{Komatsu:2010fb} 
as well as the upper bound on the spin-independent neutralino-nucleon 
scattering cross section, $\sigma^{SI}$, from XENON100 \cite{Aprile:2012nq}.
The neutralino relic density and its scattering cross section are 
computed through an interface with \textsc{micromegas}~\cite{Belanger:2006is}.
We have also modified the code \textsc{nmssmtools} to include three-body Higgs
decay in the case of the NMSSM.

We have calculated the theoretical predictions for
BR($\bmumu$) and compared them with the experimental value.
In the case of the MSSM, the SUSY contributions to this observable
\cite{Bobeth:2001jm} are likely to exceed the recent LHCb measurement
BR($\bmumu$)$=3.2{+1.5 \atop -1.2}\times 10^{-9}$ \cite{Aaij:2012nna},
especially for light pseudoscalars in the large $\tan\beta$ regime.
In the case of the NMSSM~\cite{Hiller:2004ii,Hodgkinson:2008qk},
the corresponding Wilson coefficient receives contributions from both
pseudoscalar Higgs bosons through their doublet components.
Given that the light pseudoscalar is a purely singlet-like field, it leads to
a negligible effect on this observable, and since $a^0_2$ is heavy enough
and $\tan\beta$ is small, the experimental constraint is
easily fulfilled. On the other hand, in general, BR($\bsg$) provides 
a strong constraint on SUSY models.
On top of the usual MSSM terms, NMSSM-specific contributions arise
from the extended Higgs and the neutralino sectors, which come into
effect at the two-loop level~\cite{Hiller:2004ii}.
This observable has been shown to lead to stringent constraints on the NMSSM
parameter space \cite{Cerdeno:2007sn} and this is indeed the case in our
current analysis. Here, we consider the experimental
result \cite{Amhis:2012bh} and include
the theoretical error of the calculation in the
SM \cite{Misiak:2006zs} in quadrature.

The SUSY contribution to the muon
anomalous magnetic moment, $a_\mu^{\rm SUSY}$ was also computed.
The observed discrepancy between the experimental value
\cite{Bennett:2006fi} and the SM predictions using $e^+e^-$ data favours
positive contributions from new physics in the range
$10.1\times10^{-10}<a_\mu^{\rm SUSY}<42.1\times10^{-10}$ at the $2\sigma$
confidence level \cite{Jegerlehner:2009ry,Davier:2010nc,Hagiwara:2011af}, 
combining experimental and theoretical errors in quadrature.
If tau data is employed, this discrepancy is smaller,
$2.9\times10^{-10}<a_\mu^{\rm SUSY}<36.1\times10^{-10}$ \cite{Davier:2010nc}.

We have carried out a simple scan to sample the NMSSM parameter space, 
searching for regions in the parameter space where 
Scenarios I and II could be realised. Out of the viable regions we have selected 
two benchmark points, BP1 and BP2 with parameters given in Table~\ref{table1}, 
which constitute representative examples of Scenario I and Scenario II, respectively.
The squarks and gluino masses for the first two generation are chosen to be
heavy enough to be in agreement with current LHC SUSY searches~\cite{ATLAS-susy-latest}.
We have relaxed gaugino mass unification in order to have more freedom in the 
neutralino composition, which is important in order to fix its relic 
abundance and scattering cross section. Also, trilinear parameters are non-universal 
and chosen in such a way that $a_\mu^{\rm SUSY}$ is maximized and BR$(\bsg)$ is 
in agreement with the experimental value. The resulting mass spectra are shown in
Table~\ref{table2}.


\begin{table}[t]
\begin{center}
\begin{tabular}{cc| c |c }
\hline
& Parameter & BP1 &  BP2  \\
\hline
&$\tan\beta$&5&5 \\
&$\la,\,\ka$&0.285, 0.1165& 0.286, 0.0844\\
&$A_\la,\,A_\ka$ (GeV)&670, 14.0& 820, 14.35\\
&$M_{{\widetilde L}_{i}},\,M_{{\widetilde e^c}_{i}}$
~(GeV)&300, 300& 300, 300\\
&$M_{{\widetilde Q}_{i}},\,
M_{{\widetilde u^c}_{i}},\,
M_{{\widetilde d^c}_{i}}
$~(GeV)&1000, 1000, 1000& 1000, 1000, 1000 \\
&$\mu$~(GeV)&123.5&123.5\\
&$M_1,\,M_2,\,M_3$~(GeV)&560, 1200, 1980& 240, 500, 1380\\
&$A_{\tau},\,A_{b},\,A_{t}$~(GeV)&$-1600$, 1000, 1800& $-1600$, 1000, 1300\\
 \hline
\end{tabular}
\end{center}
\caption{\label{table1}
Model parameters that define our choice of benchmark points.
The top-quark pole mass is set to 173.5 GeV and
${m_b}^{\overline{\rm MS}}(m_b)=4.18$ GeV.
}
\end{table}



\begin{table}[t]
\begin{center}
\begin{tabular}{cc| c |c }\hline
& Parameter & BP1&  BP2 \\ \hline
&$m_{\h1},\,m_{\h2},\,m_{\h3}$&97.7, 125.9, 677.0& 62.1, 125.9, 739.8 \\
&$m_{\a1},\,m_{\a2}$&3.6, 675.4& 7.6, 738.6 \\
&$m_{h^\pm},\,m_{\wt \chi^\pm_1},\,
m_{\wt \chi^\pm_2}$ &679.8, 124.0, 1192& 739, 118.1, 522.3\\
&$m_{\wt \chi^0_1},\,m_{\wt \chi^0_2},\,m_{\wt \chi^0_3}$
&87.0, 136.1, 143.3& 63.6, 124.8, 139.0\\
&$m_{\wt t_1},\,m_{\wt t_2},\,
m_{\wt b_1},\,m_{\wt b_2}$ &859.5, 1154, 1008, 1009
&951.7, 1139, 1040, 1041 \\
&$m_{\wt \tau_1},\,m_{\wt \tau_2}$&296.5, 309.5&296.5, 309.4\\
&$m_{\wt g}$&1926&1393\\
 \hline
\end{tabular}
\end{center}
\caption{\label{table2}
Relevant mass spectrum for the chosen benchmark points.
For $m_{\h2}$ the latest ATLAS limit is $125.2 \pm 0.3 {\rm(stat)}
\pm 0.6 {\rm(syst)}$ GeV~\cite{Higgs-ATLAS}. The CMS limit
corresponds to $125.3 \pm 0.4 {\rm(stat)}
\pm 0.5 {\rm(syst)}$~\cite{CMS:2012gu}. Masses are given in GeV.}
\end{table}

\begin{table}[t]
\begin{center}
\begin{tabular}{c c | c | c c}\hline
& Mass eigenstate & BP1 $(\%)$ & BP2 $(\%)$&\\ \hline
&$\h1~(S,\,H_u,\,H_d)$&79.9, 17.6, 2.5&81.2, 16.5, 2.3&\\
&$\h2~(S,\,H_u,\,H_d)$&19.5, 78.2, 2.3&18.3, 79.4, 2.3&\\
&$\h3$&95.2\% $H_d$-like&95.4\% $H_d$-like&\\
&$\a1$&99.6\% singlet-like&99.5\% singlet-like&\\
&$\a2$&95.8\% $H_d$-like&95.6\% $H_d$-like&\\
&$\wt \chi^0_1~(\wt B^0,\,\wt W^0_3,\,\wt H_d,\,\wt H_u,\,\wt S)$
&0.3, 0.2, 13.3, 30.0, 56.2&1.7, 0.9, 7.8, 22.7, 66.9&\\
&$\wt \chi^0_2$ (Higgsino, singlino)&96.8, 2.9&64.0, 29.5&\\
&$\wt \chi^0_3$ (Higgsino, singlino)&58.8, 40.9&95.6, 3.6&\\
&$\wt \chi^\pm_1$&$> 95 \%$ Higgsino-like&$> 95 \%$ Higgsino-like&\\
 \hline
\end{tabular}
\end{center}
\caption{\label{table3}
Compositions of Higges, neutralinos, and charginos for the chosen
benchmark points.
}
\end{table}

\begin{table}[t]
\begin{center}
\begin{tabular}{cc |c |c| c }\hline
& Quantity & BP1& BP2 & Range/Limit\\ \hline
&BR$(b\to s \gamma)\times 10^4$&3.86& 3.69&
2.86 -- 4.24 ($2\sigma$)\cite{Amhis:2012bh}\\
&BR$(B_s\to \mu^+ \mu^-)\times 10^9$&3.97& 3.63 &
$2.0-4.7$\cite{Aaij:2012nna}\\
&BR$(B^+\to \tau^+ \nu_\tau)\times 10^4$&1.31& 1.31 &
0.85 -- 2.89 ($2\sigma$)\cite{Lindemann:2012nn}\\
& $R_{\gamma\gamma}$
&1.0295& 1.0385
&$1.80\pm0.50$ [ATLAS]\cite{:2012gk,diphoton-ATLAS}\\
&&&&$1.48^{+0.54}_{-0.39}$ [CMS]\cite{CMS:2012gu,diphoton-CMS}\\
&
$a_\mu^{\rm SUSY}\times10^{10}$
&3.134& 6.90&
2.9 -- 36.1 ($2\sigma$)\cite{Davier:2010nc}\\
&&& &10.1 -- 42.1 ($2\sigma$) \cite{Hagiwara:2011af}\\
&Relic density ($\Omega_{\wt \chi^0_1} h^2$) &$0.082$
&0.102& 0.094 -- 0.136
[WMAP] \cite{Komatsu:2010fb}\\
&$\sigma^{\rm SI}\times10^9$ (pb)
&1.011
&0.723
&$\lesssim 1.44@86.99$ ($2\sigma$) (BP1)
\cite{Aprile:2012nq}\\
&&&&$\lesssim 1.17@63.6$ ($2\sigma$) (BP2)
\cite{Aprile:2012nq}\\
 \hline
\end{tabular}
\end{center}
\caption{\label{table4}
Low-energy observables for the chosen benchmark
point.
When scanning the parameter space,
we set $R_{\gamma\gamma} \equiv \frac{\sigma (gg\to \h2 \to \gamma\gamma)}
{\sigma (gg\to h_{SM}\to \gamma\gamma)}>0.8$
\cite{Belanger:2012tt,King:2012tr},
within the $2\sigma$ range of the ATLAS and the 
CMS result~\cite{:2012gk,CMS:2012gu,diphoton-ATLAS,diphoton-CMS}. 
For BR$(b\to s \gamma)$, 
the theoretical error is added to the experimental one in quadrature.
}
\end{table}
The compositions of Higgses, neutralinos, and charginos for our
benchmark points are shown in Table~\ref{table3}.
Since the $\mu$ parameter is chosen to be small and $\la\sim 0.3$, the singlet VEV is
around 430 GeV. Thus with the small $\kappa$ value,
the singlino mass, $2\ka v_s$ for the benchmark points BP1 and BP2 is approximately
72 and 100 GeV, respectively. Notice that this value is close to that of the 
$\mu$ parameter and that both of them are smaller than the bino and wino mass terms.
This results in an interesting hierarchical neutralino spectrum, especially 
for the three lightest neutralino states.
The
lightest neutralino is a singlino-Higgsino state (where the 
singlino component slightly dominates). There is an orthogonal eigenstate 
which is also singlino-Higgsino (but this one with a larger Higgsino component), 
which in BP1 corresponds to the third neutralino and in BP2 is the second, and 
close in mass to this one we can find an almost pure Higgsino
($\wt \chi^0_2$ for BP1 and $\wt \chi^0_3$ for BP2). 
The heavier states are bino and wino-like.

The resulting predictions for low-energy observables are shown in Table\,\ref{table4}.
Although the values of $a_\mu^{\rm SUSY}$ are in the $2\sigma$ range of the 
result from tau data, there is some tension with the one from $e^+e^-$ data. 
We have tried reducing this discrepancy by decreasing the masses in the slepton sector.
Regarding the neutralino relic abundance,
it is well known that a Higgsino-like neutralino generally entails a large
annihilation cross section and consequently a small relic abundance, but in our 
case, the presence of a sizable (slightly dominant as we said above) singlino 
component is welcome to make it compatible with the upper constraint on the dark
matter relic density. Similarly, the elastic scattering cross section for neutralinos 
with a large Higgsino component is large and can be in conflict with current
experimental bounds. In our case, the recent constraints from
XENON100 are very severe and exclude a large portion of the
parameter space we analyzed. The theoretical predictions for the spin-independent
neutralino-nucleon scattering cross section in
both BP1 and BP2 are below the current exclusion
line which, for a mass in the range 60 -- 100 GeV, is of the order of
$\lsim10^{-9}$ pb. Notice in this sense that increasing the singlino component
through the decrease, e.g., of the $\kappa$ parameter may be useful
to avoid this problem but this alternative might be questionable since
it leads to a reduction in the di-photon production from the Higgs decay.


\noindent
\section{Collider analysis}\label{results}

Let us finally
present the collider studies for the $4\ell+\MET$ signal of the benchmarks
shown in Tables~\ref{table1} and \ref{table2}.
As discussed in Sec.\,\ref{model}, the benchmark points can be
classified in terms of the Higgs masses, namely,
%
\begin{itemize}
  \item Scenario I: $(m_{h_2^0},m_{h_1^0},\,m_{a_1^0})
\equiv (126,\,98,\,3.6)$ GeV and
  \item Scenario II: $(m_{h_2^0},m_{h_1^0},\,m_{a_1^0})
\equiv (126,\,62,\,7.6)$ GeV.
\end{itemize}
In these benchmark points, BR$(h_2^0 \to a_1^0 a_1^0)$ is comparable to or larger than BR$(h_2^0 \to ZZ^\ast)$, and BR$(h_1^0 \to a_1^0 a_1^0) \sim 1$. The light pseudoscalar boson $\a1$ decays mainly into the di-tau final state with a branching fraction of $\sim$ 79 (92) \% in Scenario I (II). Given that the pseudoscalar mass is larger than 2$m_\tau$, the di-muon decay mode is negligible in Scenario II, while it remains sub-leading for Scenario I. Thus, restating the expression (\ref{signal}), the leading final state
will be
\beq\label{signal2}
  h_{1,\,2}^0 \to a_1^0 a_1^0 \to \tau^+\tau^- \tau^+\tau^-
  \to 2\ell^+ + 2\ell^- + \MET,
\eeq
with $\ell = e$, $\mu$, or $\tau_h$. Here, electrons and muons are coming from leptonic tau decays, while $\tau_h$ denotes the tau jet originated from hadronic tau decays. Occasional muons can come from $\a1\to\mu^+\mu^-$ decay process, which is sub-leading or negligible in the chosen benchmarks. The source of the missing energy is associated with neutrinos resulting from tau decays.
In order to determine the feasibility of the signal at the LHC run with the 8 TeV center-of-mass energy ($\sqrt{s}$), which has recently finished its operation, we consider inclusive search channels. In other words, our analysis differs from past studies which usually consider one or two exclusive channels such as 4$\mu$ or $2\mu + 2\tau_h$ with the 14 TeV beam condition.

We have generated Monte Carlo (MC) event samples of the Higgs signals for a proton-proton collision at $\sqrt{s}=8$ TeV using \textsc{Herwig++  2.6.1}~\cite{Bahr:2008pv} with the CTEQ6L1 parton distribution function (PDF)~\cite{Pumplin:2002vw}. $h_2^0 \to ZZ^\ast \to 4\ell$ processes are also generated since the final states are similar as in Eq.\,(\ref{signal2}).
The generated event samples have been scaled to 25 fb$^{-1}$
of integrated luminosity, which corresponds to the LHC data accumulated
in the year 2012. To simplify the analysis, we here consider leading-order
cross sections for all the processes.
The production cross section and the decay width of the SM Higgs boson
are calculated with \textsc{higlu}~\cite{Spira:1995mt}, then the
NMSSM Higgs cross sections have been obtained by corrections according
to the total decay widths and ${\rm BR}(h_{1,\,2}^0 \to gg)$ calculated
with \textsc{nmssmtools}. The generator-level events are further processed with the fast
detector simulation program \textsc{Delphes 2.0.3}~\cite{Ovyn:2009tx}
using a modified CMS detector card.
In the detector simulation, jets are reconstructed by using the
anti-$k_t$ algorithm~\cite{Cacciari:2008gp} with the radius parameter
of 0.5, and the $b$-tagging efficiency is set to be 70\%.
We assumed that the mis-tagging rate for c-jet is 10\%, and that for the gluon and light-flavor jets is 1\%, both of which are taken into account by Delphes.  
A candidate jet must satisfy $p_{\rm T} > 25$ GeV for its transverse 
momentum and $|\eta| < 2.5$ for its pseudorapidity.
Isolated electrons and muons are required to have $p_{\rm T} > 8$ and
6 GeV, respectively, and $|\eta| < 2.4$.
A tau jet is accepted only when it has $p_{\rm T} > 15$ GeV. In
order to increase the purity of the leptonic signal, any charged
lepton, including the tau jet, lying within a distance $\Delta R_{\ell j} <
0.4$ from a candidate jet is discarded in the analysis, where
$\Delta R_{ab}$ is defined as $\sqrt{(\eta_{a}-\eta_b)^2 + (\phi_{a}-\phi_b)^2}$.

\begin{figure}[t!]
  \begin{center}
    \includegraphics[width=0.48\textwidth]{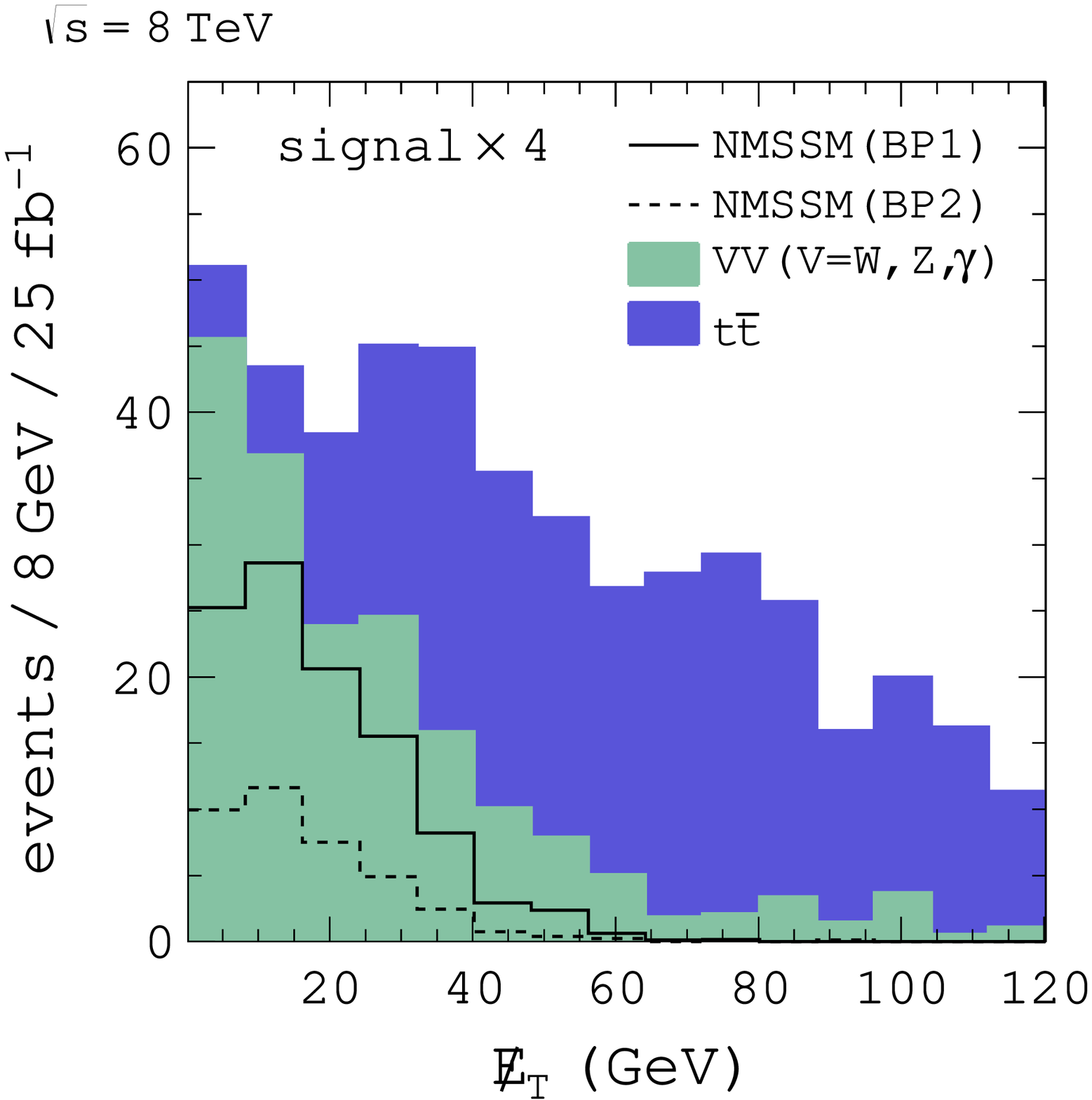}
    \includegraphics[width=0.48\textwidth]{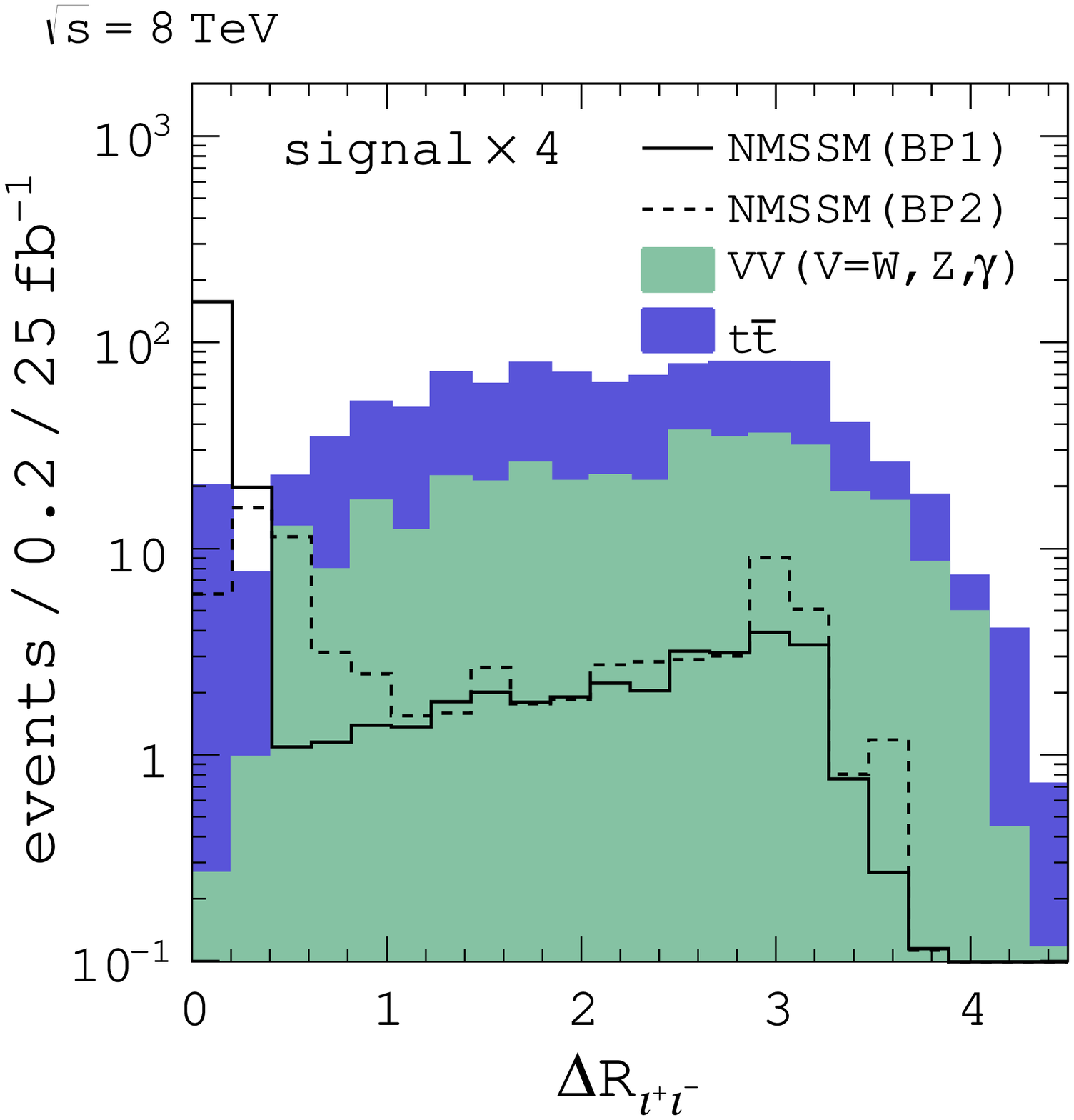}
  \end{center}
  \caption{Distributions of (left panel) the missing energy and (right
    panel) $\Delta R_{\ell^+\ell^-}$ for signals and some
    dominant backgrounds. The basic selection cuts have been
    imposed.}
  \label{fig:met}
\end{figure}

The dominant SM backgrounds include Drell-Yan (DY), $b\bar{b}$,
$c\bar{c}$, semi- or di-leptonically decaying $t\bar{t}$, $WW/Z$, and
$ZZ/\gamma$ processes.
Direct $J/\psi$ or $\Upsilon$ productions can in principle contribute,
but they are found to be almost negligible.
To estimate the backgrounds, $b\bar{b}$ and $c\bar{c}$
event samples are generated with \textsc{Pythia 6.4}~\cite{Sjostrand:2006za} 
using CTEQ6L1 PDF in the various $p_{\rm T}$ bins, and the remaining
dominant background processes are generated by \textsc{Herwig++} at
the matrix-element level. The parton showering and the hadronization
are performed by \textsc{Herwig++} for all the background processes.
Then, the MC samples have been fed into
\textsc{Delphes} using the same detector card used for the
Higgs signals. In order to suppress the backgrounds while keeping the significance of the Higgs
signals of interest, the following basic event selection
cuts are imposed.
\begin{itemize}
  \item At least two pairs of oppositely-charged leptons including the
    tau jet, 
  \item no $b$-tagged jet.
\end{itemize}
When selecting the leptons, the priority is given to the hard electrons
and isolated muons since the reconstruction efficiency of the tau jet is
relatively poor. 
No cut is applied on the missing energy, despite the presence of neutrinos
in the final state from tau decays. In fact, the neutrino momenta are 
approximately collinear with their parent tau momentum, leading to a 
partial cancellation among them and resulting in a
quite small missing energy for signal events.
To illustrate this, the signal distribution of the missing energy is compared
to some dominant background distributions in the left panel of Fig.\,\ref{fig:met}. 
As already mentioned in the previous section, multi-lepton and missing energy signatures of SUSY cascade decay processes (not included here) can also be serious backgrounds to the Higgs signal. However, the missing energy in these processes is expected to be much larger and one might attempt to use an upper cut on $\MET$. In any case, as we will see below, all these backgrounds can be efficiently suppressed by employing other cut variables.
%

\begin{figure}[t!]
  \begin{center}
    \includegraphics[width=0.48\textwidth]{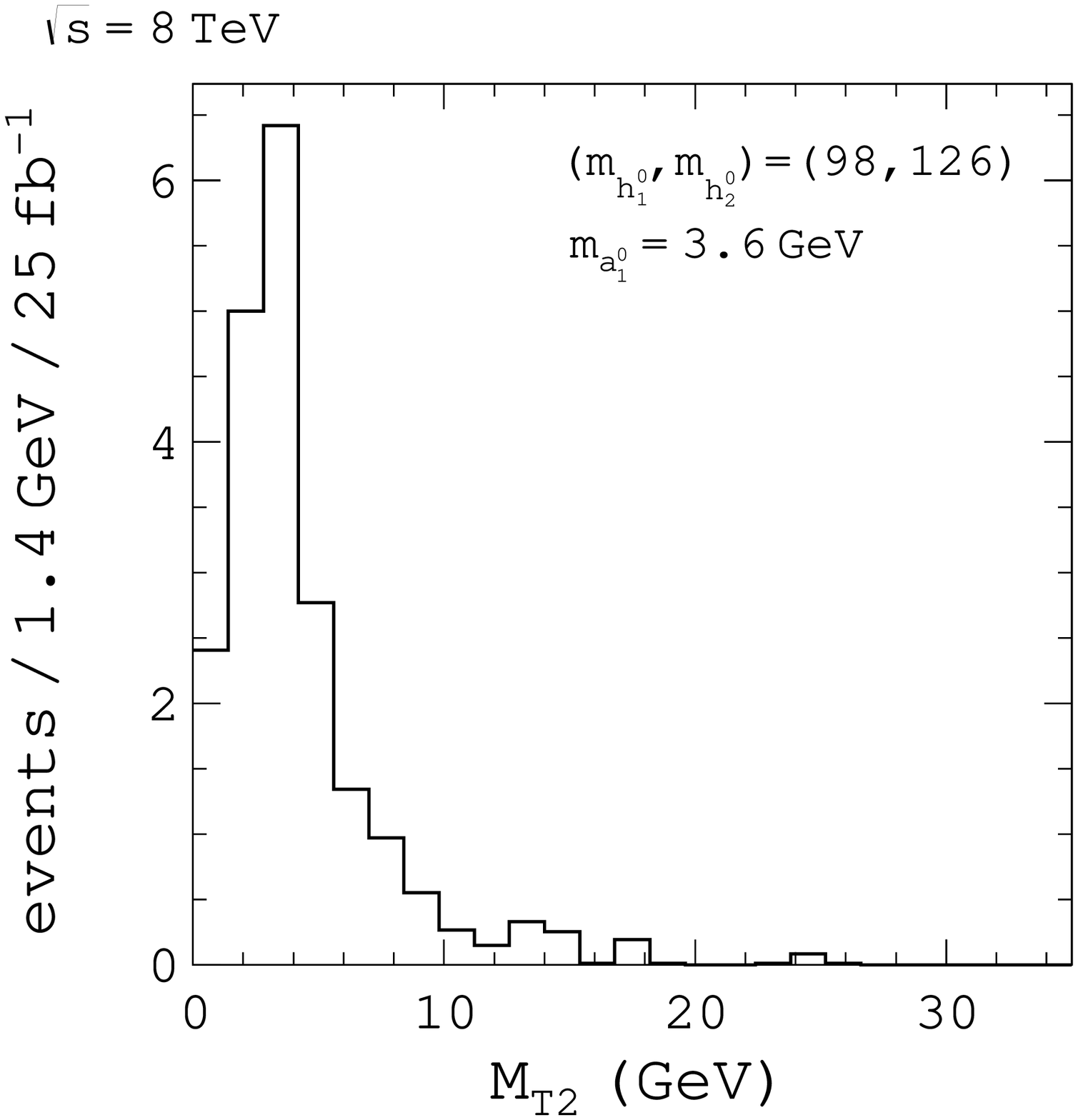}
    \includegraphics[width=0.48\textwidth]{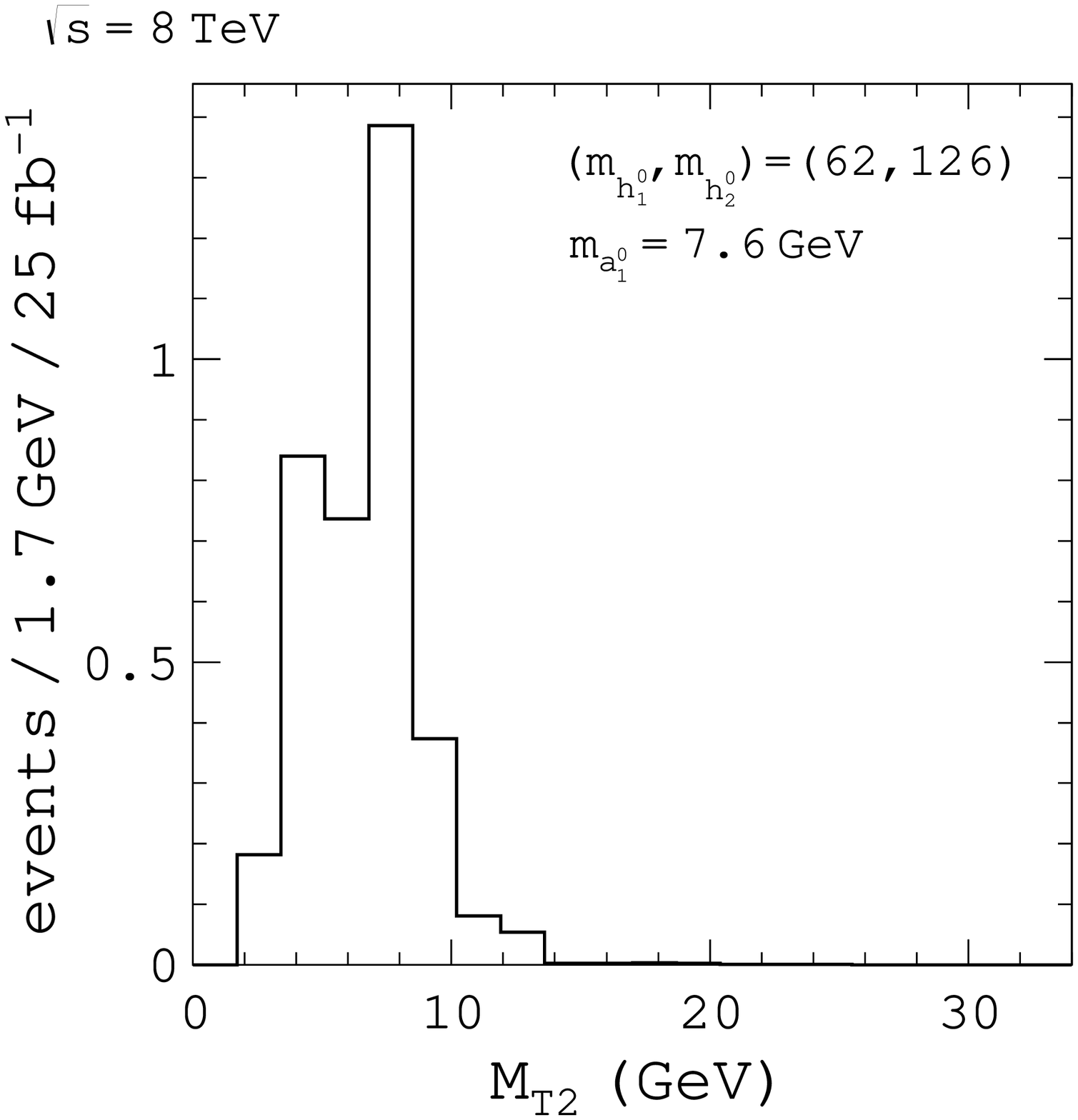}
     \end{center}
  \caption{The $M_{\rm T2}$ distributions of four visible leptons $+$
    missing energy signals for (left panel) the BP1 and (right panel) the
    BP2. All the event selection cuts described in the text except the
    jet-veto condition have been imposed.}
  \label{fig:m_T2}
\end{figure}

For the events that passed the basic selection cuts, the combinatorial
ambiguity of pairing the leptons is resolved by calculating all the
possible $M_{\rm T2}$ defined as in Ref.\,\cite{Lester:1999tx},
\beq
  M_{\rm T2} \equiv \min_{\mathbf{k}_{\rm T}^{(1)} + \mathbf{k}_{\rm T}^{(2)} =
    \ps_{\rm T}} \left[\max \left\{
    M_{\rm T}^{(1)}\left(
      \mathbf{p}_{\rm T}^{(1)},\, \mathbf{k}_{\rm T}^{(1)}
      \right),\,
    M_{\rm T}^{(2)}\left(
      \mathbf{p}_{\rm T}^{(2)},\, \mathbf{k}_{\rm T}^{(2)}
      \right)
    \right \} \right],
\eeq
where $M_{\rm T}^{(i)}$ $(i=1,\,2)$ are the transverse masses
calculated by the measured transverse momenta $\mathbf{p}_{\rm
  T}^{(i)}$ of the charged leptons and the invisible momenta
$\mathbf{k}_{\rm T}^{(i)}$ of the neutrino system, which are determined by a
numerical minimization over all possible splittings that maintain
 the missing energy condition constructed in the event.
The $M_{\rm T2}$ variable was introduced for a decay topology like a pair
production of the squark that decays into a quark and the LSP. The decay
topology given in expression (\ref{signal2}) is essentially different from that
because there are at least four neutrinos in each event. However,
the $M_{\rm T2}$ is applicable to the type of Higgs signal events
considered here since the pseudoscalar bosons are emitted back-to-back in the rest
frame of the scalar Higgs boson. Thus their decay products are
nearly collinear and the neutrinos sharing the same
parent pseudoscalar boson can be considered as one invisible particle.
In the construction of the $M_{\rm T2}$,
there are two possible ways of pairing the leptons
because two pairs of oppositely-charged leptons are initially selected.
We choose the pairing which gives the smaller value of $M_{\rm T2}$. 
This method can be justified by the fact that the $M_{\rm T2}$
value would be bounded from above by the parent particle mass $m_{a_1^0}$
if the right pairing is chosen, whereas there is no such a restriction
if the pairing was wrong~\cite{Baringer:2011nh}. Since we
here consider a light pseudoscalar boson, an upper cut on the 
$M_{\rm T2} < 25$ GeV is further implemented.
The $M_{\rm T2}$ distribution is shown in Fig.\,\ref{fig:m_T2} for
both benchmark points.
We can appreciate a very clear peak and the endpoint structure around the
pseudoscalar mass for both scenarios.

After choosing the pairing of the leptons, the transverse momentum
$p_{\rm T}^{\ell^+\ell^-} \equiv |\mathbf{p}_{\rm T}^{\ell^+} +
\mathbf{p}_{\rm T}^{\ell^-}|$ and the angular separation $\Delta
R_{\ell^+\ell^-}$ are calculated for each lepton pair.
It is likely that the leptons coming from signal events are fairly boosted in the
nearly-collinear direction. This can be observed in the right panel of
Fig.~\ref{fig:met}, and provides a clear criterion for discriminating 
the signal and background. Motivated by this, we include selection cuts on 
the $p_{\rm T}^{\ell^+\ell^-}$ and $\Delta R_{\ell^+\ell^-}$ as follows,
\begin{itemize}
  \item $p_{\rm T}^{\ell^+\ell^-} > 15$ GeV, %
  \item $\min\left\{ \Delta R_{\ell^+\ell^-}^{(1)},\,\Delta
      R_{\ell^+\ell^-}^{(2)} \right\} < 0.12$ or
    $\Delta R_{\ell^+\ell^-} < 0.65$ for both pairs.
\end{itemize}
An important observation is that after applying the above cuts, the $\h2 \to ZZ^\ast\to 2\ell^+2\ell^-$ background is found to be almost negligible since the collinearity of the leptons is not 
valid for such kind of events.
On top of that, the invariant masses of the
oppositely-charged lepton pairs are required
to be $m_{\ell^+\ell^-} < 20$ GeV to ensure that the 
lepton pairs come from the light pseudoscalar.
Collectively, a low $\Delta R_{\ell^+\ell^-}$ together
with a low $m_{\ell^+\ell^-}$ cut can discriminate
the studied signal from the $\h2 \to Z Z^\ast \to 2\ell^+2\ell^-$
process.

\begin{figure}[t!]
  \begin{center}
    \includegraphics[width=0.48\textwidth]{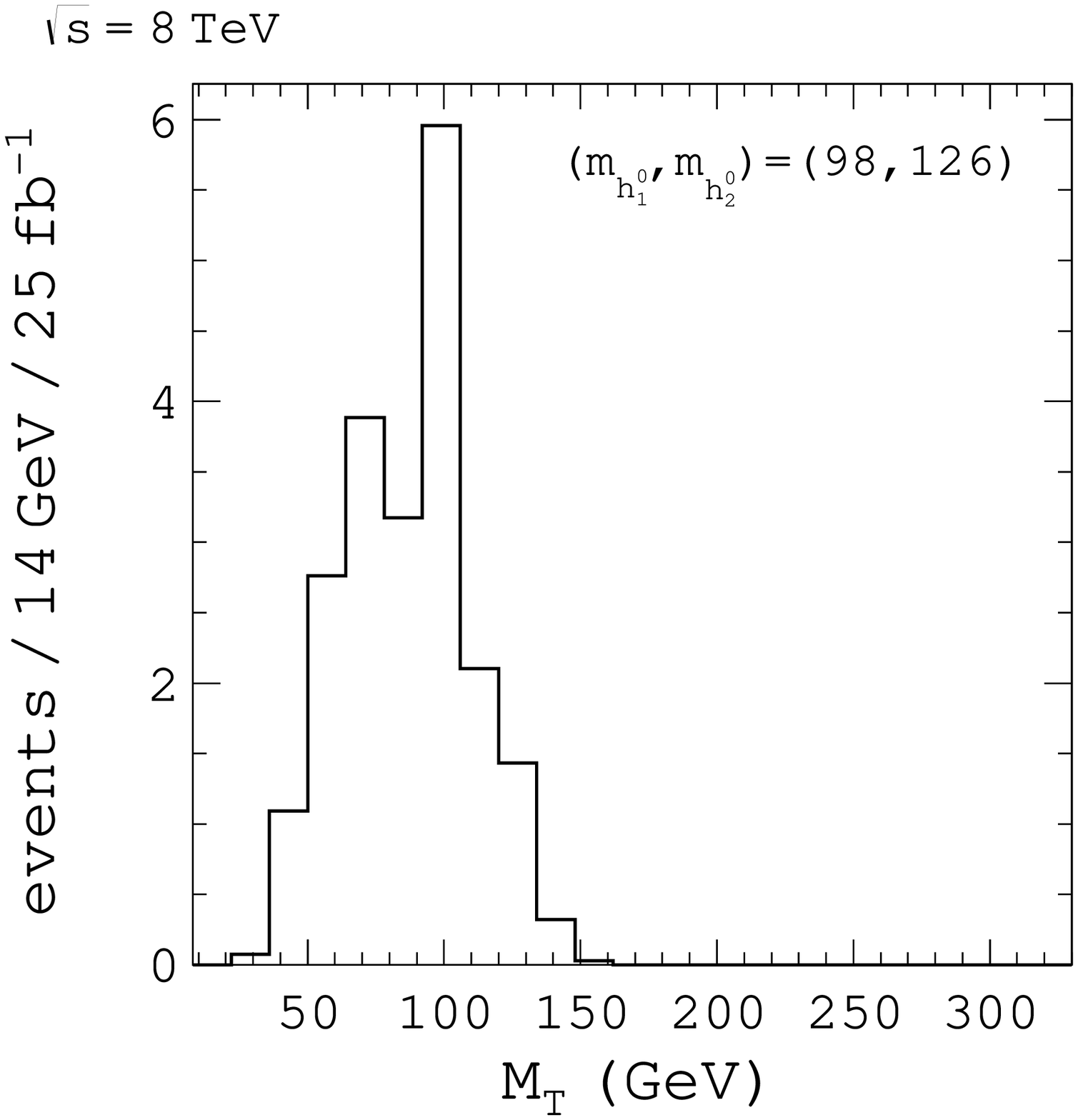}
    \includegraphics[width=0.48\textwidth]{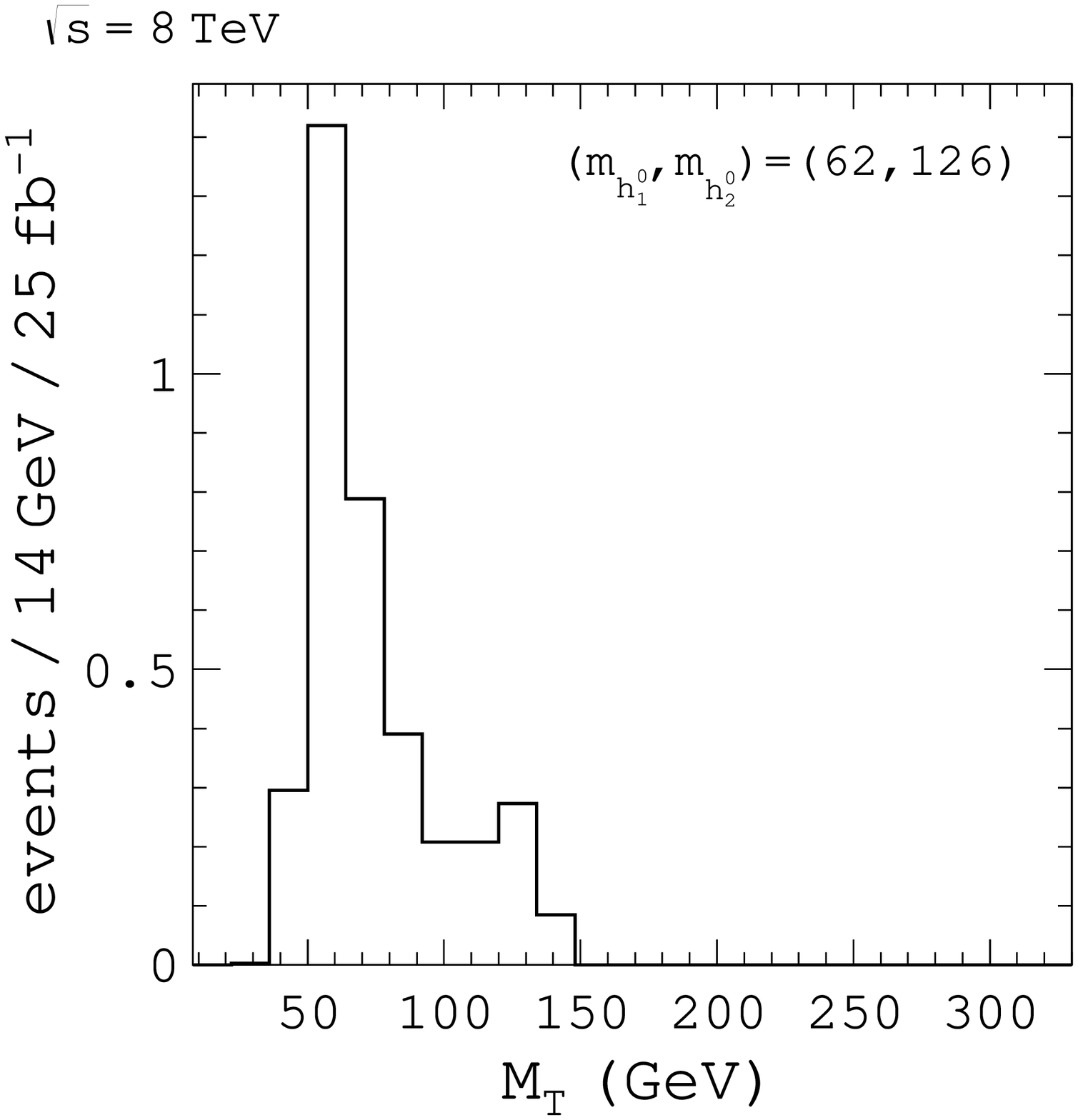}
  \end{center}
  \caption{Transverse mass distributions of four visible leptons of the
    signals for (left panel) the BP1 and (right panel) the
    BP2. All the event selection cuts described in the text except the
    jet-veto condition have been imposed.}
  \label{fig:tran_mass}
\end{figure}

We also calculate the cluster transverse mass defined as
\beq
  M_{\rm T}^2 \equiv \left(\sqrt{m_{\mathcal{V}}^2 + |\mathbf{p}_{\rm
        T}^{\mathcal{V}}|^2} +
    \MET\right)^2
    - |\mathbf{p}_{\rm T}^{\mathcal{V}} + \mathbf{\MPT}|^2,
\eeq
where $\mathcal{V}$ denotes the four-lepton system~\cite{Barger:1987re}. The
endpoint position of the $M_{\rm T}$ distribution corresponds to the
parent particle mass of the pseudoscalar bosons, i.e., $m_{h_1^0}$ or
$m_{h_2^0}$ if there exists an event with vanishing invariant mass of
the neutrinos and all the particle tracks are on the transverse plane.
However, one cannot distinguish the decay event of $h_1^0$ from
that of $h_2^0$ as they lead to the same final states.
The $M_{\rm T}$ distributions are shown in Fig.~\ref{fig:tran_mass}.
For both scenarios, the signal distributions are largely populated
around/below $m_{\h1}$, whereas  no clear peak
structure is observed around $m_{\h2}$.
The main reason is the different production cross sections of $\h1$
and $\h2$. At the generator level we obtain
$\sigma_{gg \to \h1 \to 2\a1 \to 4\ell} /
\sigma_{gg \to \h2 \to 2\a1 \to 4\ell} \sim 7$, and 
this ratio is still as large as $\sim 4$ in the detector-level
data when applying the basic selection cuts to the triggered events
in the case of the BP1\footnote{The change of the ratio is due to the 
fact that the light Higgs boson leads to relatively soft final-state 
leptons, and consequently, the acceptance at the detector becomes poor.}.
For the same reason, the $M_{\rm T}$ distribution of the BP2 has a
clear peak around the $m_{\h1}$ value as well.
Although the contribution from the
small $\h2$ decay events smears out the edge of the $M_{\rm
T}$ distribution, one can still estimate the light Higgs
boson mass scale by the peak position.
%
%
%

The $M_{\rm T}$ variable can also be used to define the signal region.
We consider the events with $35 < M_{\rm T} < 145$ GeV, which
corresponds to the region where two light Higgs bosons can be
discovered. This definition of the signal region might be useful if 
there exists a new physics process whose final state 
contains several leptons and massive invisible particles. Even 
though the new processes can contribute to the background of 
the Higgs signal, they can be readily isolated since the $M_{\rm T}$ 
values of such processes are generically large.

For events passing all the selection cuts, we also construct
the invariant mass of all visible leptons, $m_{2\ell^+ 2\ell^-}$. The
peak position of the invariant mass distribution is generically
different from $m_{\h1}$ and $m_{\h2}$ due to invisible neutrinos from
the tau decays. We impose a cut, requiring the invariant mass to be 
larger than 40 GeV, which accommodates to the possible spread of 
the invariant mass distribution.
Another possible collider variable is the
invariant mass reconstructed by the $M_{\rm T2}$-assisted on-shell
approximation of the invisible momenta~\cite{Cho:2008tj}, which has
been found to be useful for reconstructing the resonance mass peak in
the similar decay topology~\cite{Choi:2009hn}. However, by definition, 
we would need to sacrifice a certain signal statistics by imposing a 
strong $M_{\rm T2}$ cut to attain a good accuracy. 
We expect that this variable would be useful with higher luminosity. 

After imposing all the event selection cuts, 
we find that all the considered SM backgrounds turn out to be 
completely negligible.
In order to estimate the signal significance, we show how the number 
of events of the Higgs signals and the backgrounds change under each event 
selection cut in Table~\ref{table:cut}. 
The
cuts on $p_{\rm T}^{\ell^+\ell^-}$ and $\Delta R_{\ell^+\ell^-}$
are especially efficient in discriminating the signal from the
otherwise dominant DY and $b\bar b$ backgrounds, resulting in a good statistical significance for the benchmark point BP1 and a poorer result for BP2. 
We define $\hat{\mu}$ as the statistical significance calculated with the Poisson probabilities since the number of background events is small when applying all the cuts.
Indeed, the cut efficiency of each scenario is rather
different. One reason seems to be the dissimilar efficiency of the
object reconstructions.
In BP1 the decay products of the pseudoscalar boson are relatively
adjacent as it is likely that they are more boosted than in the case
of BP2, and this can affect the correct identification of the
final-state particles.

Notice finally that since there is no source of hard jets 
except for initial state radiation in the signal process, one can 
further impose a jet-veto condition that discards events containing hard jets
and can be employed for suppressing rare processes like 
$t\bar{t} +$ jets or $WW +$ jets, which have not been included in this study.
The (small) effect of this cut with 
$p_{\rm T} > 50$ GeV on the signal, is listed in the last row 
for use in studies 
that consider the other possible backgrounds.
%
\begin{table}[t!]
  \begin{center}
    \begin{tabular}{c|*{2}{c}|*{4}{c}|*{2}{l}}
      \hline&&&&&&&&\\[-2mm]
      Selection cuts & BP1 & BP2 &
      DY & $b\bar{b}$ & $t\bar{t}$ & Diboson & $\hat{\mu}_{\rm BP1}$ & $\hat{\mu}_{\rm BP2}$
      \\[2mm]
      \hline&&&&&&&&\\[-2mm]
      Basic cuts &
      26.1 & 9.5 & 252519.8 & 7998.1 & 325.2 & 186.8 & 0.05 & 0.02
      \\[2mm]
      $M_{\rm T2}$ &
      22.0 & 4.6 & ~~~1233.6 & 3225.7 & ~~~2.4 & ~~~0.5 & 0.3 & 0.07
      \\[2mm]
      $m_{\ell^+\ell^-}$ &
      21.8 & 4.3 & ~~~~145.4 & 2258.2 & ~~~1.2 & ~~~0.2 & 0.4 & 0.09
      \\[2mm]
      $p_{\rm T}^{\ell^+\ell^-}$, $\Delta R_{\ell^+\ell^-}$ &
      20.8 & 3.7 & ~~~~~~~4.7 & ~~~40.7 & ~~~0.0 & ~~~0.1 & 2.9 & 0.5
      \\[2mm]
      $M_{\rm T}$ &
      20.8 & 3.7 & ~~~~~~~4.7 & ~~~~\,0.0 & ~~~-- & ~~~0.0 & 6.7 & 1.5
      \\[2mm]
      $m_{2\ell^+ 2\ell^-}$ &
      20.1 & 3.4 & ~~~~~~~0.0 & ~~~~\,-- & ~~~-- & ~~~-- & ~-- & ~--
      \\[2mm]
      Jet veto &
      19.2 & 3.2 & ~~~~~~~--  & ~~~~\,-- & ~~~-- & ~~~-- & ~-- & ~--
      \\[2mm]
      \hline
    \end{tabular}
  \end{center}
  \caption{Cut flow of the signals and backgrounds and corresponding     
    signal significance calculated with the Poisson probabilities for
    BP1 and BP2 at the integrated luminosity of 25 fb$^{-1}$.
    See the text for the detailed descriptions of the cuts applied.}
  \label{table:cut}
\end{table}
%

To summarise, our simulation and analysis establish that the current 8 TeV data from the LHC, with dedicated selection cuts, can be used to test parts of the parameter space of non-standard Higgs scenarios with very light pseudoscalars.

\noindent
\section{Conclusions}\label{conl}

In the light of the recent detection at the LHC of a scalar boson with a mass
of 126~GeV and the possible hint for an excess in LEP that could correspond to
a second, lighter scalar we investigate the phenomenology and detectability of
viable models in the NMSSM.
We consider two scenarios that contain two light Higgses, in which the mass
of the lighter scalar is either in the range of 96 -- 100 GeV or lighter than half
of the SM-like Higgs mass. Both of them are analyzed in the presence of a very 
light pseudoscalar boson, $2 m_\tau\lsim m_{a_1^0}<2 m_b$, a range which is yet 
to be explored further at the LHC. The Higgs phenomenology of these scenarios is
extremely rich, involving cascade Higgs decays that can lead up to
eight leptons in the final state,
with or without missing energy. 
We perform a reconstruction of these channels 
for two representative benchmark points, 
determining the optimal cuts that allow us to single out the signal from the background. 
We show that the pseudoscalar
mass can be successfully determined from the $M_{\rm T2}$ distribution
of four visible leptons and the missing energy.
We also investigate the reconstruction of the light scalar Higgs 
boson from the $M_{\rm T}$ of the four visible leptons and the 
missing energy. The Higgs boson mass can be estimated by the 
peak position of the $M_{\rm T}$ distribution.
Our analysis suggests that the experimental search for inclusive 
decay modes using the current 8~TeV LHC data and dedicated cuts
can be used to test parts of the parameter space
on these possible two-Higgs scenarios with very light pseudoscalars.
%

\noindent
\subsection*{Acknowledgments}
The work of D.G.~Cerde\~{n}o and P.~Ghosh
is supported in part by the Spanish MINECO under
grants FPA2009-08958, FPA2009-09017 and FPA2012-34694, and under
the `Centro de Excelencia Severo Ochoa' Programme SEV-2012-0249,
by the Comunidad de Madrid under grant HEPHACOS S2009/ESP-1473, and by the
European Union under the Marie Curie-ITN program PITN-GA-2009-237920.
The work of C.B.~Park is supported by the CERN-Korea fellowship
through National Research Foundation of Korea.
C.~B.~Park gratefully acknowledges the hospitality of IFT UAM/CSIC,
where part of this work was carried out.

\end{document}